\documentclass[a4paper,12pt]{amsart}

\usepackage[boxruled,norelsize]{algorithm2e}

\usepackage{amssymb,amsmath,graphicx}
\usepackage{enumerate,float}
\usepackage[all]{xy}
\usepackage[margin=2cm]{geometry}

\theoremstyle{definition}
\newtheorem{thm}{Theorem}[section]

\newtheorem{prop}[thm]{Proposition}

\theoremstyle{remark}
\newtheorem{rem}[thm]{Remark}

\newcommand{\Z}{\mathbb{Z}}
\newcommand{\diag}{\mathrm{diag}}
\newcommand{\Blend}{\mathrm{Blend}}
\newcommand{\R}{\mathbb{R}}
\newcommand{\GL}{\mathrm{GL}^+}
\newcommand{\Aff}{\mathrm{Aff}^+}
\newcommand{\SO}{\mathrm{SO}}
\newcommand{\SE}{\mathrm{SE}}

\newcommand{\Sym}{\mathrm{Sym}^+}
\newcommand{\so}{\mathfrak{so}}
\newcommand{\sym}{\mathfrak{sym}}
\newcommand{\Sim}{\mathrm{Sim}^+}
\newcommand{\CO}{\mathrm{CO}^+}
\newcommand{\T}{{}^t\!}
\newcommand{\lgtr}{\mathcal{L}}
\newcommand{\sinc}{\mathrm{sinc}}

\begin{document}

\markboth{S. Kaji and H. Ochiai}
{A concise parametrisation of affine transformation}

\title{A concise parametrisation of affine transformation}
\date{}
\author{SHIZUO KAJI}
\address{Yamaguchi University / JST CREST}
\email{skaji@yamaguchi-u.ac.jp}
\author{HIROYUKI OCHIAI}
\address{Kyushu University / JST CREST}
\email{ochiai@imi.kyushu-u.ac.jp}

%


\begin{abstract}
%
Good parametrisations of affine transformations are essential to
interpolation, deformation, and analysis of shape, motion, and animation.
It has been one of the central research topics in computer graphics.
However, there is no single perfect method and 
each one has both advantages and disadvantages.
In this paper, we propose a novel parametrisation of affine transformations,
which is a generalisation to or an improvement of existing methods.
Our method adds yet another choice to the existing toolbox
and shows better performance in some applications.
A C++ implementation is available to make our framework ready to use in various applications.
\end{abstract}

\maketitle

Throughout this paper
 all vectors should be considered as real column vectors, and hence,
matrices act on them by the multiplication from the left.

\section{Introduction}
\label{sec:intro}

Affine transformation is an essential language for discussing shape and motion
(see, for example, \cite{Agoston2005}).
A common difficulty we often encounter while manipulating affine transformations
is how to represent an element.
A 3D affine transformation is represented by a $4$-dimensional homogeneous matrix (see eq. (\ref{eq:homogeneous})),
however, working directly with this representation is sometimes inconvenient since, for example;
\begin{itemize}
\item the sum of two non-singular (or rotational)
 transformations is not always non-singular (or rotational)
\item various interpolation and optimisation techniques developed for Euclidean space do not apply straightforwardly.
\end{itemize}

Many different {\em parametrisations} of transformations have been proposed
which have certain good properties. 
Nevertheless, none of them is perfect and 
we have to choose one for each purpose.
Here, we mean by parametrisation a map $\phi$ from a space
$V_G$ to the set $G$ of certain transformations.
The idea is that one can operate on and analyse transformations
in an easier space $V_G$ instead of directly dealing with the complicated set $G$.

The main goal of this paper is to propose a novel parametrisation of 3D affine transformations
which possesses the following favourable properties:
\begin{enumerate}[(I)]
\item $V_G$ is a Euclidean space
\item $\phi: V_G \to G$ is continuous and differentiable
\item $\phi$ is surjective and there exists a differentiable local inverse
$\psi: G \to V_G$ such that $\phi(\psi(A))=A$ for any $A\in G$
\item $\dim(V_G)=\dim(G)$
\item For an important subclass of transformations $H \subset G$,
there exists a linear subspace $V_H\subset V_G$ with $\phi(V_H) = H$
\item $\phi$ and $\psi$ are computationally tractable at low cost
\end{enumerate}
The condition (I) makes available the basic operations on Euclidean spaces such as 
averaging as well as interpolation, calculus,
and various linear analysis techniques
such as {Principal Component Analysis}.
Note that $\phi$ converts linear operations to non-linear ones;
addition in the parameter space $V_G$ can be highly non-linear in $G$.
The condition (II) is necessary if 
one wants to apply techniques from calculus such as differential equation
to solve optimisation problems including {\em Inverse Kinematics}.
The condition (III) means that any transformation has a canonical representative in the parameter space
and the correspondence is differentiable in both ways.
The condition (IV) means that the transformation and the parameter has 
the same degree-of-freedom and there is no redundancy.
The condition (V) means that the parametrisation restricts 
to that of a subclass.
For example, when we consider rigid transformations inside affine transformations, 
we can assure that an interpolation of rigid transformations is always rigid.
The condition (VI) is mandatory, for example, for efficient creation of deformation animation,
where millions times of computation of $\phi$ and $\psi$ per second are required.

Note that there are many good parametrisations which do not satisfy one or more 
conditions listed above but come with other nice properties.
We have to choose a parametrisation which suits a particular application,
and the purpose of this paper is to add another choice to the existing toolbox.

To clarify and demonstrate the meaning of the above conditions, we first 
look at the example of the 2D orthogonal group $\SO(2)$ 
consisting of two-dimensional rotations.
For $G=\SO(2)$, we can take $V_G$ to be the one-dimensional Euclidean space $\R$,
so that $\dim G=\dim V_G=1$.
We take $\phi$ to be the exponential map
\[
\phi(\theta)=\exp(\theta J) = \begin{pmatrix} \cos \theta &  - \sin \theta \\
\sin\theta & \cos \theta \end{pmatrix},
\]
where $J=\begin{pmatrix} 0 & -1 \\ 1 & 0 \end{pmatrix}$.
Then $\phi: V_G \rightarrow G$ gives a differentiable, surjective map,
and it is also a group homomorphism, i.e., 
$\phi(\theta_1+\theta_2) = \phi(\theta_1)\phi(\theta_2)$.
The inverse map
$\psi: G \rightarrow V_G$ is given by 
\[
\psi \left(\begin{pmatrix} a & -b \\ b & a \end{pmatrix} \right) = 
\begin{cases} 
\tan^{-1}\frac{b}{a} + k\pi & (a \neq 0) \\
 \pi/2 + k\pi & (a=0, b= \pm 1) \\
 \end{cases},
\]
where $-\pi/2 < \tan^{-1}\frac{b}{a} < \pi/2$ is the principal value of the arctangent and $k\in \Z$.
Note that $\psi$ cannot be taken as a globally continuous map on $G$,
but locally around any point of $G$ it can be taken to be a differentiable map by choosing $k\in \Z$ appropriately.
Therefore, in this situation, the conditions (I), (II), (III), and (IV) are satisfied.
The condition (V) is out of question here since there is no interesting subclass 
of $\SO(2)$.
As for (VI), the computational cost of $\phi$ and $\psi$ is reasonable since
we have the closed formulae as described above.
When $G$ is a more complicated class, 
we cannot expect this kind of simple solution.

The mathematical tool we employ 
to construct our parametrisation
 is {\em Lie theory}, particularly, 
the Lie group-Lie algebra correspondence and 
the {\em Cartan decomposition} of Lie algebras 
(for mathematical background, we refer the reader to textbooks on Lie theory 
such as \cite{Knapp}).
We also discuss several applications of our parametrisation.
This paper is an improved and extended version of \cite{meis2013}.
Introductory explanations of our method including the background
can be found in 
\cite{AO,Ochiai:2013:MDM:2542266.2542268,Ochiai:2014:MBM:2614028.2615386}.

\section{Related Work}
\label{sec:related}
Precedent researches have given 
a number of different parametrisations to the subclasses of 
3D affine transformations
listed in Figure~\ref{fig:transformations} (see also \S \ref{sec:comparison}).
One of the earliest example is the parametrisation of the group of 3D rotations $\SO(3)$
 by the Euler angle. As is well-known, it suffers from the gimbal lock.
 In our language, the condition (III) is not satisfied.
The gimbal lock issue is avoided when $\SO(3)$ is parametrised with the invertible quaternions.
However, the dimension of the space of invertible quaternions is $4$, which 
is greater by one than that of $\SO(3)$. 
This means we need an extra variable and the condition (IV) is not satisfied.
The parameter space is almost Euclidean; 
except that the origin $0$ of the parameter space
does not correspond to any transformation. Although this is only a single point, it can be problematic. For example,
we have to be careful when interpolate transformations so that the interpolation curve does not pass the point. 
If we restrict ourselves to the unit quaternions, we have the parameter space of the same dimension as $\SO(3)$.
However, the parameter space is non-Euclidean 
and special techniques are required for basic operations
(see, for example, \cite{Shoemake1985,Barr1992,Nielson1993,Kim1995,Ramamoorthi1997,Buss2001}).

\begin{figure}
\label{fig:transformations}
\[
\xymatrix@=20pt{ 
  \Aff(3)\ar@{-}[d]\ar@{-}[rd] & & \\
\Sim(3) \ar@{-}[d] \ar@{-}[rd] &  \GL(3)\ar@{-}[rd]\ar@{-}[d] \\
\SE(3)\ar@{-}[rd]\ar@{-}[d] &  \CO(3) \ar@{-}[d] \ar@{-}[rd]& \Sym(3)\ar@{-}[d] \\
\R^3  & \SO(3) &  \R^+
}
\]
\begin{tabular}{|c|l|}
\hline
symbol & transformation \\
\hline
$\R^3$ & the group of translations \\
$\SO(3)$ & the group of rotations 
 (or the {\em special orthogonal group})\\
$\R^+$ & the group of dilations \\
$\SE(3)$ & the group of rigid transformations
  (or the group of {\em Euclidean motions} or {\em screw motions}) \\
$\CO(3)$ & the group of linear conformal transformations with positive determinants \\
$\Sym(3)$ & the set of scale-shear transformations (or the set of positive definite symmetric matrices)\\
$\Sim(3)$ & the group of similarity transformations with positive determinants \\
$\GL(3)$ & the group of linear transformations with positive determinants \\
$\Aff(3)$ & the group of affine transformations with positive determinants \\
\hline
\end{tabular}
\caption{Hierarchy of 3D transformations. The upper classes contain the lower classes.}
\end{figure}

The group of rigid transformations $\SE(3)$ can be parametrised with the dual quaternions,
which is a generalisation of the parametrisation of $\SO(3)$ by the quaternions.
In \cite{DQ}, they used the dual quaternions to blend 
rigid transformations and their method works particularly well in skinning.
However, for other applications we may be troubled by the complicated structure of the parameter space;
the space of the unit dual quaternions is the semi-direct product of the group of the unit quaternions
 and $\R^3$.

Let $\Aff(3)$ be the group of 3D affine transformations with positive determinants.
Note that an (invertible) affine transformation has the positive determinant 
if and only if it contains no flip (or reflection).
In \cite{Shoemake1992} a method 
to interpolate elements of $\Aff(3)$ using the polar decomposition
was introduced.
Transformations are decomposed into the rotation, the scale-shear,
and the translation parts,
and then SLERP was used for interpolating the rotation part 
and the linear interpolation was used for the rest.
This idea to look at $\Aff(3)$ as the (non-direct) product of 
three spaces $\SO(3), \Sym(3)$, and $\R^3$ 
has been fundamental and many of the current graphics systems adopt it.
However, the parametrisations of $\SO(3)$ by the quaternions and 
$\Sym(3)$ by matrices are not Euclidean. 

On the other hand, in \cite{Alexa2002}
a definition of
scalar multiple and addition in $\Aff(3)$ is given based on
the idea to parametrise $\Aff(3)$ by the corresponding Lie algebra, which gives a Euclidean parameter space.
This is a generalisation of \cite{Grassia98}, where $\SO(3)$ is parametrised by its Lie algebra.
The same idea is also used in \cite{SAM}.
A notable feature of their construction, which is missing in \cite{Shoemake1992}, is that the scalar multiplication satisfies
``associativity.''
That is, for $\alpha, \beta\in \R$,
the $\alpha$-multiple of the $\beta$-multiple of a transformation is
equal to the $\alpha\beta$-multiple of it. 
However, a major defect of their construction is that it 
does not work with transformations
 with negative real eigenvalues. 
 That is, there is no representatives for some transformations
 and the condition (III) does not hold.
 This causes a big problem (see \S \ref{sec:efficiency}).
 Our work can be considered as a workaround of this inconvenience
 at the cost of loosing the associativity.
 In addition, our parametrisation comes with a few advantages including
 a fast closed formula and better handling of large rotation.
    

\section{The parametrisation map}
\label{sec:results}
The key idea of our method is to linearise the curved space of the Lie group of 
the affine transformations using the Lie algebra.
This direction has already been pursued in the computer graphics community, for example, in \cite{Grassia98,Alexa2002,SAM}.
However, we have to be careful about the following points when we deal with the group $\Aff(3)$;
\begin{itemize}
\item the correspondence between the group $\Aff(3)$ and its Lie algebra is not one-to-one.
\item working in the Lie algebra is not intuitive since the meaning of each parameter is unclear
(for example, finding consistent 
logarithm discussed in \S \ref{sec:rot} becomes difficult).
\item it involves 
the high cost computations of the matrix exponential and logarithm.
\end{itemize}

To avoid those shortcomings, we propose a handy and mathematically rigorous
parametrisation.
Using the Cartan decomposition of the Lie algebra of $\GL(n)$,
we establish a parametrisation of $\Aff(3)$ which satisfies all the conditions listed in the previous section.
(For (V), we can take all the important subclasses listed in the previous section as $H$.)

Recall that $\Aff(3)$ denotes the group of $3$-dimensional affine transformations 
with positive determinants, i.e., the connected component including the identity.
In other words, $\Aff(3)$ is the group of 3D orientation preserving affine transformations.
As usual, we represent elements in $\Aff(3)$ by $4\times 4$-homogeneous matrices.
\begin{equation}\label{eq:homogeneous}
 \Aff(3)= \left\{ A=
 \begin{pmatrix} a_{11} & a_{12} & a_{13} & l_x\\
a_{21} & a_{22} & a_{23} & l_y \\
a_{31} & a_{32} & a_{33} & l_z\\
 0 & 0 & 0 & 1 \end{pmatrix} \mid \det(A)>0
 \right\}
\end{equation}
We often denote by $A_{ij}$ the $(i,j)$-entry of $A$.
From this representation, it is clear that $\Aff(3)$ is a $12$-dimensional Lie group.
We call the upper-left $3\times 3$ part of $A\in \Aff(3)$ 
as the {\em linear part} of $A$ and denote it by $\hat{A}$,
and we denote by $\psi_L(A)$ the {\em translation part} of $A$
\[
\psi_L(A) := 
 \begin{pmatrix} 1 & 0 & 0 & l_x\\
0 & 1 & 0 & l_y \\
0 & 0 & 1 & l_z\\
 0 & 0 & 0 & 1 \end{pmatrix}. 
 \]
Let $M(N)$ be the set of $N \times N$-matrices.
Denote by $\iota$ the standard inclusion $M(3) \to M(4)$
given by
\[
 \iota(B)=\begin{pmatrix} B & 0\\
  0 & 1 
 \end{pmatrix}.
\]
Then, $A=\psi_L(A)\iota(\hat{A})$ for $A\in \Aff(3)$.

We define our $12$-dimensional parameter space as the product of two $3$-dimensional and one
 $6$-dimensional vector spaces:
\[
 V_{\Aff(3)} := \R^3 \times \so(3) \times \sym(3),
\]
where 
\[
 \so(3) := \left\{ X\in M(3) \mid X=-\T X \right\}
\]
is the set of the $3\times 3$-anti symmetric matrices (which is the Lie algebra of $\SO(3)$) and
\[
 \sym(3):=
 \left\{ Y\in M(3) \mid Y=\T Y \right\}
\]
is the set of the $3\times 3$-symmetric matrices.
To sum up, our parameter space $V_{\Aff(3)}$ is the $12$-dimensional Euclidean space
of the form
\[
 \R^3 \times \so(3) \times \sym(3) = \left\{ \left(
 \begin{pmatrix} 
 1 & 0 & 0 & l_1\\
 0 & 1 & 0 & l_2 \\
 0 & 0 & 1 & l_3 \\
 0& 0& 0& 1
 \end{pmatrix},
 \begin{pmatrix} 
 0 & x_4 & x_5 \\
 -x_4 & 0 & x_6 \\
 -x_5 & -x_6 &  0
 \end{pmatrix},
 \begin{pmatrix} 
 y_7 & y_8 & y_9 \\
 y_8 & y_{10} & y_{11} \\
 y_9 & y_{11} & y_{12} 
 \end{pmatrix} \right)
  \right\},
\]
where we identified $\R^3$ with the translation matrices
so that the identity matrix serves as the origin.
We emphasise that there is no restriction on the parameters
and each variable can take any real number.

The {\em parametrisation map} is defined by
\begin{eqnarray}\label{exp}
\phi:  \R^3 \times \so(3) \times \sym(3) &\to& \Aff(3)  \\
 (L, X, Y) &\mapsto& L \cdot \iota(\exp(X)\exp(Y)),   \nonumber
\end{eqnarray}
where $\exp$ is the matrix exponential.
A locally differentiable inverse
$\psi$ is defined as follows:
\begin{eqnarray}\label{log}
\psi=(\psi_L,\psi_R,\psi_S):  \Aff(3) &\to& \R^3 \times \so(3) \times \sym(3) \\
 A &\mapsto& (\psi_L(A),\log(\hat{A} S^{-1}),\log(S)),
\nonumber
\end{eqnarray}
where $S=\sqrt{\T\hat{A}\hat{A}}$.
That is, for any $A\in \Aff(3)$ 
we have $\psi(A) \in V_{\Aff(3)}$  
(unique up to modulo $2\pi$, as we see later in \S \ref{sec:rot})
such that
\[
 \phi(\psi(A))=A.
\]
It is obvious from the definition that our parametrisation satisfies the conditions (I), (II), and (IV) in \S 1.
For (V), we set
\begin{align*}
V_{\Sim(3)} &= \{ (L,X,cI_3)\in V_{\Aff(3)} \mid c\in \R \} \\
V_{\GL(3)} &= \{ (I_4,X,Y)\in V_{\Aff(3)} \} \\
V_{\SE(3)} &= \{ (L,X,0)\in V_{\Aff(3)}  \} \\
V_{\CO(3)} &= \{ (I_4,X,cI_3)\in V_{\Aff(3)}  \mid c\in \R\} \\
V_{\Sym(3)} &= \{ (I_4,0,Y)\in V_{\Aff(3)}  \} \\
V_{\R^3} &= \{ (L,0,0)\in V_{\Aff(3)} \} \\
V_{\SO(3)} &= \{ (I_4,X,0)\in V_{\Aff(3)} \} \\
V_{\R^+} &= \{ (I_4,0,cI_3)\in V_{\Aff(3)}  \mid c\in \R\},
\end{align*}
where $I_3, I_4$ are the identity matrices.
When restricted to the above subspaces, our parametrisaion
gives those for the corresponding subclasses of the transformations.
We give explicit fomulae of $\phi$ and $\psi$ in the next section,
and see that the conditions (III) and (VI) are satisfied as well.

\begin{rem}
The product $\exp(X)\exp(Y)$ in \eqref{exp} is in fact the polar decomposition,
where $\exp(X)\in \SO(3)$ is the rotation part and $\exp(Y)\in \Sym(3)$ is the scale-shear part.
Therefore, there are a few choices for the order of the product in \eqref{exp}.
(Then, \eqref{log} would change accordingly.)
However, we believe the current order is the most intuitive when
the user works directly in the parameter space.
Usually, we are comfortable to think of a rigid transformation
as a motion so that it is specified with the global frame.
On the other hand, scaling and shearing ``changes'' the shape of an object 
so it is more natural to think that it takes place with the local frame.
Therefore, we would like to first perform scaling and shearing, which is parametrised by $\sym(3)$,
  so that the standard frame of $\R^3$ and the local frame are aligned.
Then, the rigid transformation, parametrised by $\R^3\times \so(3)$, is applied.
\end{rem}

\section{Computing the parametrisation map}\label{sec:parametrisation}
In this section, we give closed formulae
for the parametrisation maps
(\ref{exp}) and (\ref{log}).
For a C++ implementation, we refer the reader to \cite{affinelib}.

The matrix exponential is defined by the infinite series as in
Appendix \ref{ap:exponential}.
Since computation by the infinite series is very slow, 
it is crucial to have an efficient algorithm
for applications.
We give fast closed formulae 
including ones for the exponential and the logarithm of symmetric matrices,
which have their own interests.
\subsection{Closed formula for (\ref{exp})}\label{closed-formula-of-exp}
First, we look at the definition (\ref{exp}).
The closed formulae for the exponential maps are given as follows.
For $\so(3)$ part, Rodrigues' formula \cite{rodrigues} computes:
\begin{equation}\label{rodrigues}
 \exp(X)=I_3+\dfrac{\sin\theta}{\theta}{X}+\dfrac{1-\cos\theta}{\theta^2}{X}^2
 =I_3+\sinc(\theta) X + \frac12 \left(\sinc\frac{\theta}{2} \right)^2 X^2,
\qquad \theta=\sqrt{\dfrac{\operatorname{\mbox{tr}}(\T{X} {X})}{2}}.
\end{equation}
When $|\theta|$ is very small, 
the following second order approximation can be used to avoid small denominator:
\begin{equation}\label{sinc}
 \sinc(\theta) \approx 1-\dfrac{\theta^2}{6}.
\end{equation}

For $Y\in \sym(3)$, the computation of $\exp(Y)$ can be done by diagonalisation
(see \eqref{eq:diag} in the appendix).
However, we introduce a faster method avoiding the high-cost computation of diagonalisation 
 (see Appendix \ref{ap:computation-of-exp} for the derivation):
Observe that the eigenvalues
$\lambda_1, \lambda_2, \lambda_3$ are the roots of the characteristic polynomial 
and obtained by the cubic formula. 
This means, $\lambda_i$'s can be computed by a closed formula and the computational cost is much cheaper 
than that of diagonalisation.
Assume that $\lambda_1 \ge \lambda_2 \ge \lambda_3$.
and put $Z=Y-\lambda_2 I_3$. Then, the eigenvalues of $Z$ are
$\lambda'_1= \lambda_1-\lambda_2, \lambda'_2=0, \lambda'_3=\lambda_3-\lambda_2$.
Put
\begin{align*}
b &= 1 - \dfrac{\lambda'_1 \lambda'_3 (e_2(\lambda'_1)-e_2(\lambda'_3)) }{\lambda'_1-\lambda'_3} \\
c &= \dfrac12 + \frac{\lambda'_1 (2e_2(\lambda'_1)-1)-\lambda'_3 (2e_2(\lambda'_3)-1)}{2(\lambda'_1-\lambda'_3)},
\end{align*}
where $e_2(x)=\dfrac{\exp(x)-1-x}{x^2}$.
Then, we have
\begin{equation}\label{formula-exp}
 \exp(Y) = \exp(\lambda_2)(I_3 + bZ + cZ^2).
\end{equation}
We may have small denominators when some of $\lambda_i$ collide, and thus, $\lambda'_1\to 0$ and/or $\lambda'_3\to 0$. 
Nevertheless, the above formula converges.
In practice, when $x$ is small we can use the second order approximation
\[
 e_2(\lambda'_i) \approx \frac12 + \frac{\lambda'_i}{6} + \frac{{\lambda'}_i^2}{24}.
\]
\subsection{Closed formula for (\ref{log})}\label{closed-formula-of-log}
To compute \eqref{log}, we provide formulae for
$\log(S)$ and $\log(R)$, where $S=\sqrt{\T\hat{A}\hat{A}}$ and $R=\hat{A} S^{-1}$.

First, note that $\T\hat{A}\hat{A}$ is symmetric positive definite, and hence,
the square root and its logarithm are uniquely determined.
We can calculate them by the diagonalisation as well, 
however, we adapt a similar method to the one given in the previous subsection.
Denote the eigenvalues of $\T\hat{A}\hat{A}$ by 
$\lambda_1, \lambda_2, \lambda_3 > 0$.
Assume that $\lambda_1 \ge \lambda_2 \ge \lambda_3$ and
set $Z=\T\hat{A}\hat{A}/\lambda_2$. The eigenvalues of $Z$ are $\lambda'_1=\lambda_1/\lambda_2, 1 ,\lambda'_3=\lambda_3/\lambda_2$.
Define
\begin{align*}
a &= -1 + \dfrac{\lambda'_3 \lgtr_2(\lambda'_1) - \lambda'_1 \lgtr_2(\lambda'_3)}{\lambda'_1-\lambda'_3} \\
c &= \dfrac{\lgtr_2(\lambda'_1)-\lgtr_2(\lambda'_3)}{\lambda'_1-\lambda'_3},
\end{align*}
where $\lgtr_2(x)=\dfrac{\log(x)-(x-1)}{x-1}$.
Then, we have
\begin{equation}\label{formula-log}
 \log(S)=\frac12 \log(\T\hat{A}\hat{A})
  = \frac12 \left((a+\log(\lambda_2))I_3 - (a+c)Z + cZ^2 \right).
\end{equation}
When $\lambda'_i\to 1$, we use the second order 
approximation
$\lgtr_2(\lambda'_i) \approx -\frac{\lambda'_i-1}{2} + \frac{(\lambda'_i-1)^2}{3}$.
To obtain $R$, we compute $S^{-1} = \exp( -\log(S) )$ by 
the formula (\ref{formula-exp}).
Since the eigenvalues of $-\log(S)$ are
$-\log(\lambda_i)/2$, we can reuse the values of $\lambda_i$.

For $R \in \SO(3)$, by inverting Rodrigues' formula we have
\begin{equation}\label{log-SO(3)}
 \log(R)= \dfrac{1}{2\sinc\theta}({R}-\T {R}), \quad
\theta=\cos^{-1}\left(\dfrac{\mathrm{Tr}({R})-1}{2}\right),
\end{equation}
where $\cos^{-1}$ takes the principal value in $[0, \pi]$.
When $\theta$ is small, the approximation \eqref{sinc} can be used.
When $\pi-\theta$ is small, there is a different method.
Let $v=(v_1,v_2,v_3)$ be the eigenvector of $R$ with eigenvalue one.
Then $\log(R)=\theta\begin{pmatrix} 0 & -v_3 & v_2 \\ v_3 & 0 & -v_1 \\ -v_2 & v_1 & 0 \end{pmatrix}$,
where $\theta=\epsilon\cos^{-1}\left(\dfrac{\mathrm{Tr}({R})-1}{2}\right)$ with
 $\epsilon=\begin{cases} 1 & (v_2({R}-\T {R})_{13}\ge 0) \\ -1 & (v_2({R}-\T {R})_{13}<0) \end{cases}$.
Alternatively, one can look at the diagonal entries in \eqref{rodrigues}
to compute $\log(R)$ (see \cite[Appendix A.1.1]{Pennec1997} for details).

As we mentioned before, here we have indeterminacy of $\cos^{-1}$ up to modulo $2\pi$.
This reflects the fact that our parametrisation can handle (or distinguish) rotations
by more than $2\pi$. 
If we impose continuity, we will have one explicit choice as 
is described in the next subsection.

\subsection{Indeterminacy up to $2\pi$}\label{sec:rot}
The definition of (\ref{log}) contains an ambiguity up to 
a factor of $2\pi$.
In this section, we see how we give an explicit choice 
and how it is useful in some cases.

Let $R=\hat{A} S^{-1} \in \SO(3)$, which
appears in the definition (\ref{log}).
One could simply take the principal value of the logarithm,
 i.e., choose $0\le \theta < \pi$ in \eqref{log-SO(3)}.
On the other hand, it is sometimes convenient to
 distinguish the $2\pi$-rotation around the $x$-axis from the identity,
and also from $2\pi$-rotation around the $y$-axis.
For example, if we consider a rotational motion rather than the final position of it,
we have to track the rotational degree larger than $2\pi$. 
(see Figure~\ref{fig:over360degree}, where the target shape is twisted over $2 \pi$ degrees).

In such a case, we can choose a particular value for the logarithm 
using the coherency or
continuity regarding the metric on the parameter space.
Algorithm \ref{alg:continuous-log} computes 
the logarithm of $R$ which is closest to a given $X'$.
 
\begin{algorithm}[htb]\label{alg:continuous-log}
  \SetAlgoLined
 \KwIn{rotation matrix $R\in \SO(3)$, anti-symmetric matrix 
 $X'\in\so(3)$}
 \KwOut{logarithm of $R$ closest to $X'$}
\Begin{
 $\theta \gets \cos^{-1}((\mathrm{Tr}({R})-1)/2)$ \\
 \If{$\theta=\pi$}{
 \eIf{$\sqrt{\operatorname{\mbox{tr}}(\T{X'} {X'})/2}>0$}
 	{\Return{$\dfrac{\pi}{\sqrt{\operatorname{\mbox{tr}}(\T{X'} {X'})/2}} X'$}}
    {\Return{$\begin{pmatrix} 0 & \pi & 0\\ -\pi & 0 & 0\\ 0 & 0& 0\end{pmatrix}$}}
    }
 $X \gets \dfrac{1}{2\sin\theta} ({R}-\T{R})$ \\
 $\theta' \gets \sqrt{{X'}_{12}^2+{X'}_{13}^2+{X'}_{23}^2} $ 
 \If{${X}_{12}{X'}_{12}+{X}_{13}{X'}_{13}+{X}_{23}{X'}_{23}<0$}{
 $X \gets -X$, $\theta \gets -\theta$
 }
 \While{$\theta'-\theta>\pi$}{
  $\theta \leftarrow \theta+2\pi$
  }
 \While{$\theta-\theta'>\pi$}{
 $\theta \leftarrow \theta-2\pi$
 }
 \Return{$\theta {X}$}
}
\caption{consistent logarithm of rotation matrices}
\end{algorithm}

\subsection{Computational Efficiency}\label{sec:efficiency}
We compare our closed formulae for exponential and logarithm given in \S \ref{sec:parametrisation}
with two widely used algorithms;
Pad\'e approximation and the scaling-and-squaring method by Higham (\cite{Higham2008,Higham09thescaling})
implemented in Eigen library (\cite[version 3.2.8]{eigenweb}), and diagonalisation \eqref{eq:diag}.
Our implementation is given in \cite{affinelib}.
We computed the matrix exponential and logarithm
for randomly generated $10^6$ matrices for $10$ times and measured the average computational time.
The timing is given in Table~\ref{table:timing}.
Error is measured by the maximum of the squared Frobenius norm $|X-\exp(\log(X))|^2_F$
among all the $10^6$ matrices.

\begin{table}[H]\small
\caption{Timing: measured with 1.7Ghz Intel Core i7, 8GB memory, single thread}
\begin{tabular}{c|ccc}
 & $\exp$ of $\sym(3)$ &  $\log$ of $\Sym(3)$ & Error in $\Sym(3)$
 \\ \hline
 Ours &  0.2402s & 0.2685s & $4.792 \times 10^{-27}$ \\
 Higham & 0.5454s & 5.930s & $6.356\times 10^{-29}$  \\
 Diag & 0.5720s  & 0.6195s & $3.419\times 10^{-23}$
\end{tabular}
\label{table:timing}
\end{table}

We also compare different parametrisation of $\Aff(3)$.
Alexa/SAM directly computes $\log$ and $\exp$
for $\Aff(3)$ (suggested in \cite{Alexa2002,SAM}).
Polar+DQ computes the polar decomposition and then use
dual quaternions for the rotation and the translation parts (suggested in \cite{DQ}),
and positive definite symmetric matrices for the scale-shear part.
Polar+Grassia also computes the polar decomposition but uses $\log$ for the rotation part
(suggested in \cite{Sumner2005}).
For computing the polar decomposition, 
we used the algorithm in \cite{Higham86}, which is more efficient than by SVD and 
widely used in the computer graphics community (see \cite{Shoemake1992}).
Error is measured by the maximum of the squared Frobenius norm $|X-\psi(\phi(X))|^2_F$
among randomly generated $10^6$ matrices with $\det(X)>10^{-3}$.
We noticed that for $X$ with a very small determinant, the error tends to get bigger
with our method.
Hence, to deal with near singular matrices, it is better to use Higham's algorithm to 
obtain the polar decomposition and then use our formulae for $\exp$ and $\log$ to 
compute the parametrisation.
\begin{table}[H]\scriptsize
{\caption{Comparison of various parametrisations}}
\begin{tabular}{c|ccccccc}
 & Surjectivity & Extrapolation & Associativity & Large rotation & $\phi$ & $\psi$ & Error\\ \hline
Ours & Yes & non-singular & No & Yes & $0.4835s$ & $0.3543s$ & $1.936\times 10^{-25}$ \\
Alexa/SAM & No & can be singular & Yes & No & $12.88$ & $0.8432$ & $3.269 \times 10^1$ \\
Polar+DQ & Yes & can be singular & No & No & $0.5366$ & $0.03531$ &  $7.025\times 10^{-28}$\\
Polar+Grassia & Yes & can be singular & No & Yes & $0.5777$ & $0.1053$ & $7.371\times 10^{-28}$
\end{tabular}
\label{table:comparison}
\end{table}
Surjectivity means that any element $X\in \Aff(3)$ can be parametrised.
Extrapolation indicates whether extrapolation of two elements with any weights can fall out of $\Aff(3)$ or not.
Surjectivity and extrapolation affects robustness of the method.
Associativity means the associativity of scalar multiple explained in \S \ref{sec:related}.
Large rotation indicates the capability of dealing with rotations greater than $2\pi$
(see \S \ref{sec:rot}).

Notice that due to the lack of surjectivity, Alexa/SAM has a huge error 
(see \S \ref{sec:comparison} for theoretical background).


\section{Application}
\label{sec:example}

In this section, we will give a few applications
 of our parametrisation with comparison with existing techniques
(see also \cite{tetrisation} for a further comparison).
Source codes with the MIT licence are available at 
\verb+https://github.com/shizuo-kaji/+.

\subsection{Blending transformations}
\label{blend}
One direct application of Euclidean parametrisation of transformation in general is
blending different transformations (see, for example, \cite{Arsigny2003}).
Suppose that transformations $\{ A_i \mid 1\le i \le n \}$ and
the corresponding weights $\{ w_i\in \R \mid 1\le i \le n\}$ are given.
One way to blend them is 
to take the linear sum in the parameter space:
\begin{equation}\label{eq:blend}
 \Blend( w_1,\ldots,w_n, A_1,\ldots, A_n ) := \phi \left( \sum_{i=1}^n w_i \psi(A_i) \right).
\end{equation}
Note that the above formula reduces to interpolation of 
two transformations when $n=2$ and $w_2 = 1-w_1$.
One can also apply standard techniques such as B-spline 
for interpolating three or more transformations.

With the condition (V),
if $A_i$'s are all in the same class,
the blended transformation \eqref{eq:blend} always stays
in the same class regardless of the weights.

\subsection{Shape deformer}
\begin{figure}
\begin{center}
   \includegraphics[width=2.6cm,bb=0 0 240 240,clip]{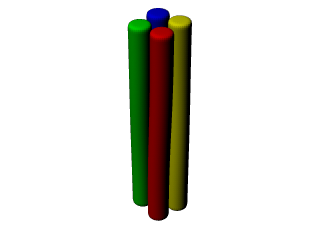}
   \includegraphics[width=2.6cm,bb=0 0 240 240,clip]{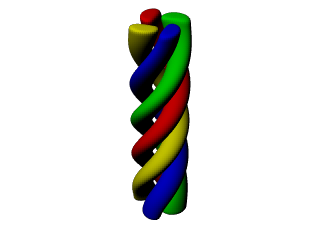}
   \includegraphics[width=2.6cm,bb=0 0 240 240,clip]{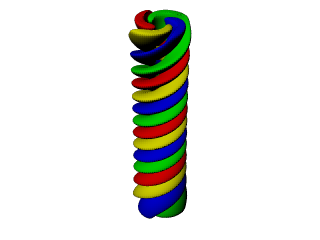}
   \includegraphics[width=2.6cm,bb=0 0 500 410,clip]{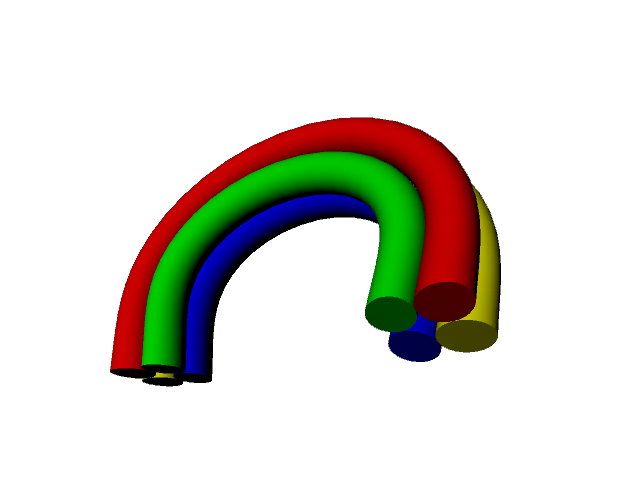} \\
   \includegraphics[height=3.3cm]{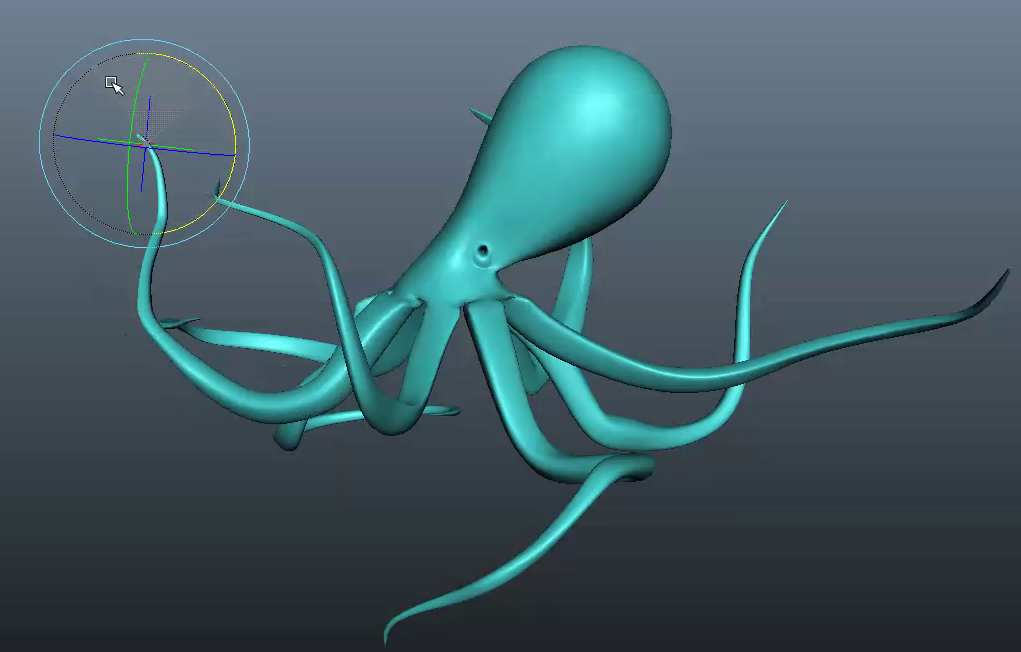}   
   \includegraphics[height=3.3cm]{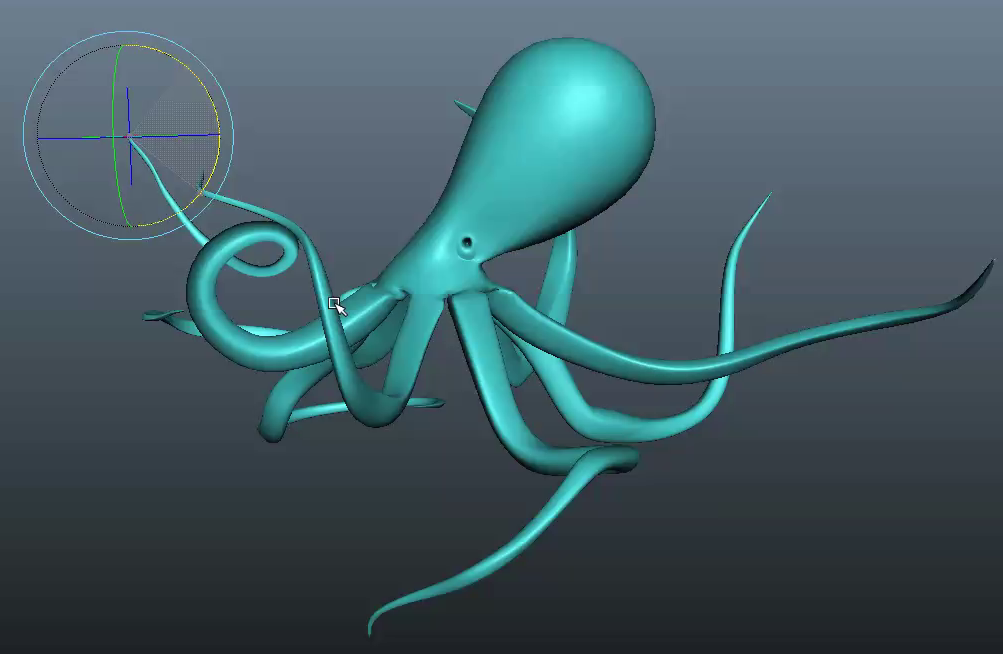}
   \includegraphics[height=3.3cm]{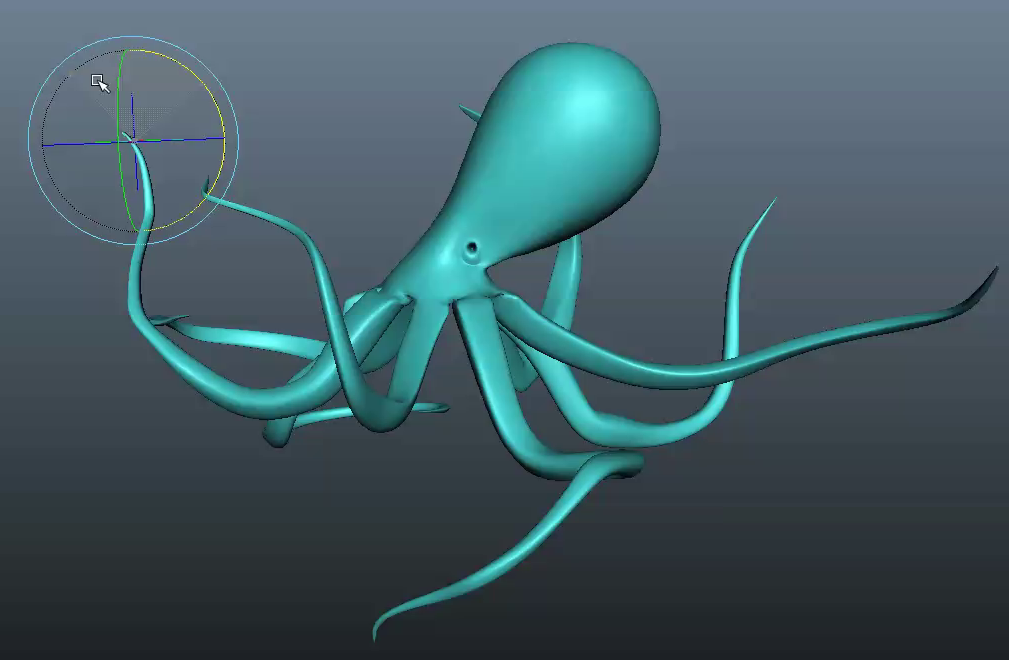}   
  \caption{Upper row: 
  The four strings (left) are being deformed by our method.
  Two probes are placed above and below the strings, one fixed to the identity and the other rotated and translated.
  Notice that more than $2\pi$ rotation is recognised.
  Lower row: The leg of the octopus model (left) is being deformed by a probe rotated by large degrees.
  With our method (middle) the leg is made swirl, while with quaternions (right) one sees almost no deformation.
  }
  \label{fig:over360degree}
\end{center}
\end{figure}

As a simple application of the $\Blend$ function, we construct a shape deformer 
(see Figure~\ref{fig:over360degree}) 
similar to the one developed in \cite{Llamas2003}. 
A much more elaborated version is discussed in \cite{probe} using our parametrisation,
so we just give a simple idea of it here.
 
Assume that a set of affine transformations $\{A_i\in \Aff(3) \mid 1\le i\le n\}$
(specified by ``probe'' handles) is given.
We want to deform a mesh with vertices $U$, according to a given weight function
$w_i: U \to \R \quad (1\le i \le n)$.
The weight can be painted manually or calculated automatically
from the distance of the vertex and the probe.

Using the blending function defined in \S \ref{blend},
we compute the deformed position of $u\in U$ by
\[
 u \mapsto \Blend(w_1(u),\ldots,w_n(u), A_1,\ldots, A_n ) \cdot u.
\]
See \cite{probe} for the details.

\subsection{Shape blender}\label{ex:nway}
The authors considered in \cite{SCA12}
to apply our parametrisation to morph two isomorphic 2D meshes
based on the idea in \cite{Alexa2000}.
It was extended to a shape blending algorithm for an arbitrary number of 3D meshes
in \cite{tetrisation}. We briefly review a simple version of it.

Suppose that we are given a rest mesh $U_0$ and 
target meshes $U_i \quad (1\le i\le n)$.
We assume that all the meshes are compatibly triangulated, i.e.,
a one-to-one correspondence between triangles of $U_0$ and 
each $U_i$ is given.
We want to produce a blended shape $U(w_1,\ldots, w_n)$
with respect to the user specified weights $\{ w_i\in \R \mid 1\le i\le n\}$.
We require that $U(w_1,\ldots, w_n)$ interpolates the given shapes, more precisely, 
$U(w_1,\ldots, w_n)=U_0$ if $w_i=0 \ (1\le \forall i\le n)$, and
$U(w_1,\ldots, w_n)=U_k$ if $w_i=\begin{cases} 1 & (i=k) \\ 0 & (i\neq k) \end{cases}$. 

\begin{figure}
\noindent\begin{minipage}[b]{.57\textwidth}
  \includegraphics[width=.7\textwidth]{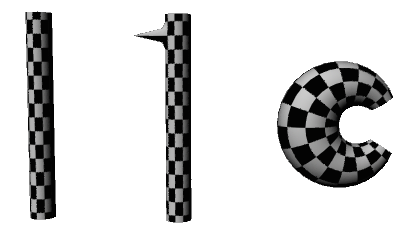}\\
  \includegraphics[width=.7\textwidth]{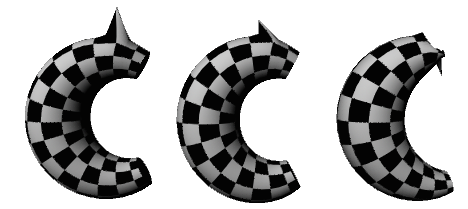} \\
 Upper row left to right: rest shape $U_0$ and target meshes $U_1,U_2$.
 Lower row: shapes are blended with $w_1=w_2=0.8$
 by our method (left), by \cite{Alexa2002,SAM} (middle),
 and by polar decomposition and quaternions \cite{Shoemake1992} (right).
 The left picture looks the most natural.  The thorn in the middle picture is slanted because 
 \cite{Alexa2002,SAM} mixes up rotation and shear parts.
 In the right picture, the thorn totally collapse because quaternions cannot handle large rotation.
\end{minipage} 
\hfill
\begin{minipage}[b]{.4\textwidth}
  \includegraphics[width=.99\textwidth]{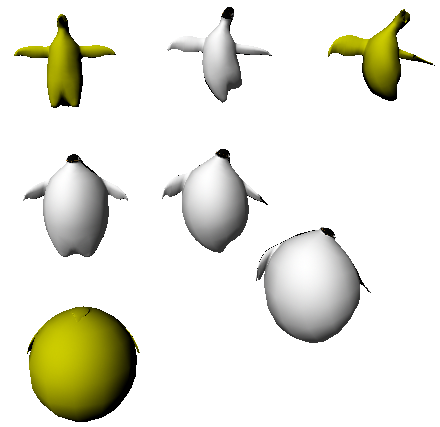}\\
{Three yellow penguin shapes are blended with our method to produce variations (white).}
\end{minipage}
 \caption{Shape blending}
\label{fig:nway}
\end{figure}

First, we associate for each face $f_{ij} \quad (1\le j \le m)$ of $U_i$ 
the unique affine transformation 
$A_{ij}$ which maps the corresponding face $f_{0j}$  
to $f_{ij}$ and 
the unit normal vector of $f_{0j}$ to that of $f_{ij}$.
Then, we put
\[
A'_j(w_1,\ldots, w_n):= \Blend(w_1,\ldots,w_n,A_{1j},\ldots,A_{nj}) \in \Aff(3).
\]
Note that those blended transformations
are not coherent on the edges 
so we cannot apply them directly to $U_0$.
In order to obtain a blended shape,
we have to ``patch'' $A'_j$'s to 
obtain a set of affine transformations consistent on the edges of $U_0$, that is,
 a piecewise linear transformation.
This is done by finding the minimiser of the error function 
$\sum_{j=1}^m |A_j(w_1,\ldots, w_n)-A'_j(w_1,\ldots, w_n)|^2$
to obtain a piecewise linear transformation
$A_j(w_1,\ldots, w_n) \quad (1\le j\le m)$.
A blended shape is then computed by
\[
U(w_1,\ldots, w_n)= \bigcup_{1\le j\le m}  A_{j}(w_1,\ldots, w_n) f_{0j}
\]
(see \cite{tetrisation} for the detail).
This is useful, for instance, to produce variations of shapes
from a given set of examples (see Figure \ref{fig:nway}).

\subsection{Pose interpolation}\label{ex:pose}
\begin{figure}
 \begin{center}
  \includegraphics[width=.32\textwidth]{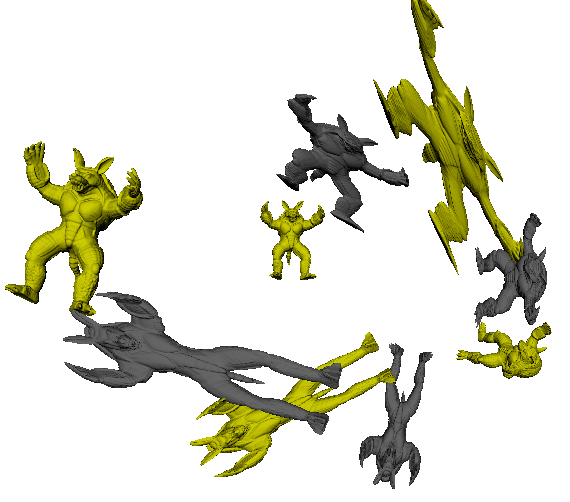} \ 
  \includegraphics[width=.32\textwidth]{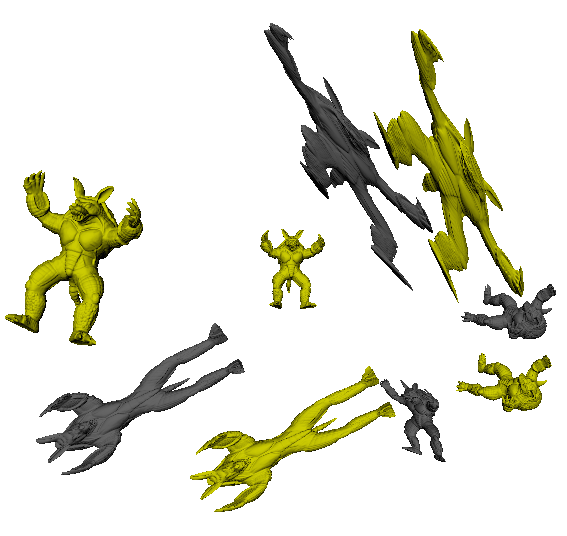} \ 
  \includegraphics[width=.32\textwidth]{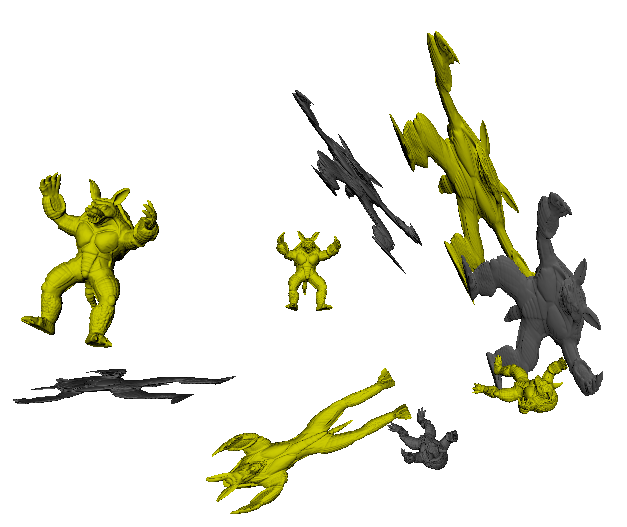}
   \end{center}
 \caption{Pose Interpolation: red rectangles are interpolated by the B-spline curve in the parameter space.
 Left to Right: our method, matrix logarithm \cite{Alexa2002}, and homogeneous matrices (linear). 
 The linear method shows degeneracy. 
 }
\label{fig:pose}
\end{figure}

Our method also has a direct application to interpolating transformations
(cf. \cite{Kim1995,Kim95,SAM}).
Given a set of key frames for 
positions and orientations of objects such as cameras or joints.
That is, we are given for each object,
a sequence of affine transformations $A_{j} \quad (1\le j \le m)$.
We want to interpolate the key frames to produce an animated poses
(see Figure~\ref{fig:pose}).
Since our parametrisation takes value in the $12$-dimensional Euclidean space,
any ordinary interpolation method such as Hermite polynomial,
 B\'ezier, and B-spline can be applied in the
parameter space: Let 
$
\mathrm{Interpolate}(t, \{\mathrm{knots}\})
$
be an interpolation curve with a given set of knots in $\R^{12}$.
 Then, we can compute interpolated poses by
\[
 A(t):=\phi(\mathrm{Interpolate}(t, \{ \psi(A_{j}) \mid 1\le j \le m\})).
\] 
With the condition (V),
if $A_j$'s are all in the same class $H$,
the interpolated transformation 
always stays in the same class $H$
regardless of the value of the time parameter $t$.

\section{Discussions: Comparison to precursors}\label{sec:comparison}
In this section, we discuss the relationship between 
our method and previous techniques from a theoretical point of view.
Existing parametrisations of transformation are classified roughly into three types:
based on matrix, Clifford algebra, and Lie algebra.

Using matrix, we can parametrise $\Aff(3)$ by $4\times 4$-matrices as in 
(\ref{eq:homogeneous}).
It is easy to work with and computationally efficient.
Also it is equipped with the additive structure 
(in addition to the multiplicative structure)
so that transformations can be blended by taking the weighted sum.
However, the main drawback is that there are matrices which do not correspond
to elements in $\Aff(3)$, namely those with vanishing determinant.
This is problematic in many applications.
For example, addition and scalar multiplication are not closed
in $\Aff(3)$. If you simply take the weighted sum of matrices,
you may get degenerate transformation.
Furthermore, the operations do not restrict to the important subclasses.
For example, the sum of two rotational matrices is not rotational.

Quaternions (respectively, dual quaternions (\cite{DQ}))
can be used to parametrise $\SO(3)$ (respectively, $\SE(3)$).
More generally, using Clifford algebras
one can parametrise various kinds of transformations.
For example, the anti-commutative dual complex numbers (\cite{DCN})
parametrise $\SE(2)$ and conformal geometric algebra (CGA, for short)
can be used to present $\Sim(3)$.
In fact, CGA deals with a larger class
of transformation including non-linear ones
(\cite{Hestenes,CGA2001,Vince2008,Hildenbrand04geometricalgebra,Wareham2004,Dorst2007,Wareham2008}).
Clifford algebra based parametrisations are usually very efficient
and the weighted sum can be used to blend transformations.
However, there are two main disadvantages:
first, it requires special construction for each class of transformation
and, in particular, there are no known construction for 
parametrising the entire $\Aff(3)$.
Among those listed in Figure~\ref{fig:transformations}, constructions for
$\R^3, \SO(3),\SE(3)$, and $\Sim(3)$ are known.
The other disadvantage is that the parameter space is not Euclidean
and one cannot directly apply techniques from linear analysis and
calculus. In addition, the parameter space  generally has
greater dimension than the group of transformations
to be parametrised.

To cover the entire $\Aff(3)$,
Shoemake (\cite{Shoemake1992})
suggested to first decompose an element of $\Aff(3)$ 
into the rotation, the scale-shear, and the translation parts by the polar decomposition
and then parametrise rotational part by quaternions and the rest by matrices.
This is one of the most commonly used method.
However, the parameter space is not Euclidean; 
the rotational part is parametrised by the unit quaternions 
and the scale-shear part
by the positive definite symmetric matrices.
None of these parameter spaces is Euclidean
and it causes problems in some applications
(see Figure \ref{fig:over360degree} and \ref{fig:nway}).

In mathematics and physics, it is well-known that
Lie algebras are useful to parametrise Lie groups.
The exponential map is a differentiable map
from the Lie algebra $\mathfrak{g}$ to the Lie group $G$.
When $G$ is compact, 
this is surjective and gives a parametrisation of $G$ by 
a Euclidean space $\mathfrak{g}$ of the same dimension as $G$,
which satisfies (II) and (IV) discussed in \S \ref{sec:intro}.
For a Lie subgroup $H$ of $G$, there exists a Lie sub-algebra
$\mathfrak{h}$ and the image of the exponential map is contained in $H$.
Hence, (V) is also satisfied.
Grassia (\cite{Grassia98}) 
introduced the machinery
to computer graphics community
by suggesting to use the Lie algebra $\so(3)$ to parametrise $\SO(3)$.
In \cite{Sumner2005}, they combined this idea with the polar decomposition 
to blend elements in $\GL(3)$.
They parametrised the rotation part by $\so(3)$ and 
the scale-shear part by $\Sym(3)$.

In \cite{Alexa2002,SAM}, they used the Lie algebra
to parametrise the whole $\Aff(3)$.
However, $\Aff(3)$ is not compact and
the exponential map is not surjective ((III) fails to be satisfied).
There exist transformations 
which are not parametrised;
$C\in \GL(n)$ is in the image of the exponential map 
if and only if $C=B^2$ for some $B$. In other words, $\log(A)$ for $A\in \Aff(3)$
exists only when $\hat{A}=B^2$ for some $B$ (see \cite{Culver66}, for detail).
Although those transformations are not so many as was discussed in \cite{SAM},
in practice, it is problematic if one has to always take care of those exceptions
(see Table \ref{table:comparison}) and it does affect the result in some applications (see Figure \ref{fig:nway}).
Another disadvantage is the cost of computation
since the computation
of the exponential and the logarithm usually involves iteration
(see Table \ref{table:comparison}).

Our construction is also based on the Lie algebra.
To remedy the problem mentioned above we uses the Cartan decomposition in the Lie algebra,
 which corresponds to the polar decomposition in the Lie group.
 More precisely, for $A\in \Aff(3)$, $\exp(\psi_R(A)) \exp(\psi_S(A))$ is the polar decomposition of 
 $\hat{A}$. 
This enables us to achieve (III) while keeping the advantages of Lie algebra
based methods.
Note that thanks to our closed formulae for the exponential and the logarithm
needed for our parametrisation, 
we avoid the cost inefficient computation of the polar decomposition.
In fact, the above relation provides an alternative method 
to compute the polar decomposition.

There are two main drawbacks of our method.
First, with our parametrisation, the $\Blend$ function defined in \S \ref{blend}
fails to be {\em bi-invariant}, that is, invariant under the left and the right translation, 
whereas DLB developed in \cite{DQ} for $\SE(3)$
has this property (see also \cite{Park}, for interpolation).
Bi-invariance is very important in some applications
such as skinning. In \cite{arsigny:inria-00071383}, 
a bi-invariant mean of elements in a connected Lie group in general is studied.
However, computing bi-invariant means and interpolation is very costly,
and as far as the authors are aware, there is no parametrisation for $\Aff(3)$
which achieves bi-invariance with a simple blend function.
(In \cite{Alexa2002}, they mistakenly claim that their method achieves bi-invariance.)
Second, our $\Blend$ function lacks in associativity, which is satisfied by 
\cite{Alexa2002,SAM}.

\section{Conclusions and future work}
\label{sec:conclusion}
We proposed a concise $12$-dimensional Euclidean parametrisation for 
three-dimensional affine transformations.
Our parametrisation has some good properties compared to
the existing ones; most notably,
it parametrises the entire $\Aff(3)$ by the 
 Euclidean space of the same dimension
 and it handles transformations containing large rotation.
The Euclidean nature allows us to apply various
techniques developed for the Euclidean space.
The computational cost is relatively cheap and a C++ implementation is provided.
On the other hand, scalar multiple and addition in the parameter space 
do not map nicely to the transformation space;
a straight line in the parameter space does not map to a geodesic in the transformation space.

Since standard calculus and linear algebra are valid on our parameter space,
we can apply differential equations, linear data analysis, 
and optimisation techniques on the group of affine transformations.
Possible applications include
 inverse kinematics (\cite{Sumner2005}), 
 filtering captured data (\cite{Lee2002}),
 motion analysis and synthesis (\cite{TWCAR09}), 
 and interpolation (\cite{Park}).
 We will discuss them elsewhere.
 
 It would also be interesting to extend our result to parametrise
wider classes of transformations such as the projective transformations.
\appendix
\section{Exponential and logarithm maps}
\label{ap:exponential}
We recall the definition of the matrix exponential and the logarithm maps.
For a square matrix $A$,
the exponential map is defined by
\begin{equation}\label{ap:def-of-exp}
 \exp(A)=\sum_{i=0}^{\infty} \dfrac{A^i}{i!}.
\end{equation}
The ``inverse'' of the exponential is called the logarithm.
It should satisfy the equation $\exp(\log(X))=X$.
Let $f$ denote $\exp$ or $\log$.
We have
$f(PAP^{-1})=Pf(A)P^{-1}$ for any invertible matrix $P$.
Hence, when $A$ is diagonalisable, that is, 
there is an invertible matrix $P$ such that
$P^{-1}AP=\diag(d_1,d_2,\ldots,d_n)$,
we can compute
\begin{equation}\label{eq:diag}
\begin{aligned}
 f(A)&=f(P\diag(d_1,d_2,\ldots,d_n)P^{-1})\\
 &=Pf(\diag(d_1,d_2,\ldots,d_n))P^{-1} \\
 &=P \diag(f(d_1),f(d_2),\ldots,f(d_n)) P^{-1}.
\end{aligned}
\end{equation}
Note that $\log(X)$ does not always exist and even when it exists,
 it is not necessarily unique in general (\cite{Culver66}).

\section{Derivation of the exponential and logarithm formulae in \S \ref{sec:parametrisation}}
\label{ap:computation-of-exp}

For $Y\in \sym(3)$
we consider $f(Y)$, where $f=\exp$.
(And respectively, for $Y\in \Sym(3)$, we consider $f(Y)$, where $f=\log$.)
The main idea  is to divide the infinite series (\ref{ap:def-of-exp}) by
the degree three characteristic polynomial to reduce it to a degree two polynomial (see, for example, \cite{Moler78}).
More explicitly, let $p_Y(x)=\det(xI_3-Y)$ be the characteristic polynomial of $Y$.
Then by dividing by the degree three polynomial $p_Y(x)$, we have
\[
f(x) = q(x)p_Y(x) + r(x),
\]
where 
$r(x)=a+bx+cx^2$ is the reminder.
By Cayley-Hamilton's theorem, we know $p_Y(Y)=0$, so 
\begin{equation}\label{eq:remainder}
f(Y) = r(Y)=aI_3+bY+cY^2.
\end{equation}
Since $Y$ is symmetric,
we can diagnalise it to have 
$Y = P D P^{-1}$, where $D=\diag(\lambda_1,\lambda_2,\lambda_3)$.
To determine $a,b,c \in \R$, consider
$f(D)=\diag(f(\lambda_1), f(\lambda_2), f(\lambda_3))$
and 
\begin{align*}
f(D) &=P^{-1} f(Y) P = P^{-1} (aI_3+bY+cY^2) P \\ &= aI_3 + bD + c D^2.
\end{align*}
Then, $a,b$, and $c$ are obtained by solving 
the following Vandermonde's linear system
\[
\begin{pmatrix}
1 & \lambda_1 & \lambda_1^2 \\
1 & \lambda_2 & \lambda_2^2 \\
1 & \lambda_3 & \lambda_3^2 \\
\end{pmatrix}
\begin{pmatrix}
a \\ b \\ c 
\end{pmatrix}
=
\begin{pmatrix}
f(\lambda_1) \\ f(\lambda_2) \\ f(\lambda_3) 
\end{pmatrix}.
\]
\begin{prop}
When the eigenvalues are pairwise distinct,
the coefficients in \eqref{eq:remainder} is given by
\begin{equation}\label{eq:exp-abc}
\begin{aligned}
s &= f(\lambda_1)/((\lambda_1-\lambda_2)(\lambda_1-\lambda_3)), \\
t &= f(\lambda_2)/((\lambda_2-\lambda_3)(\lambda_2-\lambda_1)), \\
u &= f(\lambda_3)/((\lambda_3-\lambda_1)(\lambda_3-\lambda_2)), \\
a &= s\lambda_2\lambda_3 + t\lambda_3\lambda_1 + u\lambda_1\lambda_2, \\
b &= -s(\lambda_2+\lambda_3) -t(\lambda_3+\lambda_1) - u(\lambda_1+\lambda_2), \\
c &= s+t+u.
\end{aligned}
\end{equation}
\end{prop}
Next, we make this formula more robust
so that it is valid even when some of the eigenvalues coincide.
Assume $\lambda_1 \ge \lambda_2 \ge \lambda_3$ and 
put $\bar{Y}=Y-\lambda_2 I_3$ and $\bar{f}(x)=f(x+\lambda_2)$. 
By this substitution, the computation of $f(Y)$ is reduced to that of $\bar{f}(\bar{Y})$,
where the eigenvalues of $\bar{Y}$ are 
$\bar{\lambda}_1=\lambda_1-\lambda_2,\bar{\lambda}_2=\lambda_2-\lambda_2=0,
\bar{\lambda}_3=\lambda_3-\lambda_2$.
Denote by $T_2(x)$ the analytic function $\dfrac{\bar{f}(x)-\bar{f}(0)-\bar{f}'(0)x}{x^2}$.
By \eqref{eq:exp-abc}, the coefficients in $\bar{f}(\bar{Y})=\bar{a}I_3+\bar{b}\bar{Y}+\bar{c}\bar{Y}^2$ are computed as
\begin{eqnarray*}
\bar{s} &=& \bar{f}(\bar{\lambda}_1)/(\bar{\lambda}_1(\bar{\lambda}_1-\bar{\lambda}_3)), \\
\bar{t} &=& \bar{f}(0)/(\bar{\lambda}_3\bar{\lambda}_1), \\
\bar{u} &=& \bar{f}(\bar{\lambda}_3)/((\bar{\lambda}_3-\bar{\lambda}_1)\bar{\lambda}_3), \\
\bar{a} &=& \bar{t}\bar{\lambda}_3\bar{\lambda}_1 =\bar{f}(0), \\
\bar{b} &=& -\bar{s}\bar{\lambda}_3 -\bar{t}(\bar{\lambda}_3+\bar{\lambda}_1) - \bar{u}\bar{\lambda}_1 \\
&=&  \frac{1}{\bar{\lambda}_1\bar{\lambda}_3(\bar{\lambda}_1-\bar{\lambda}_3)} 
\{ -\bar{f}(\bar{\lambda}_1) {\bar{\lambda}_3}^2-\bar{f}(0)(\bar{\lambda}_1-\bar{\lambda}_3)(\bar{\lambda}_3+\bar{\lambda}_1) + \bar{f}(\bar{\lambda}_3){\bar{\lambda}_1}^2 \} \\
&=&  \frac{1}{\bar{\lambda}_1\bar{\lambda}_3(\bar{\lambda}_1-\bar{\lambda}_3)} 
\{ -(\bar{f}(\bar{\lambda}_1)-\bar{f}(0)) {\bar{\lambda}_3}^2 + (\bar{f}(\bar{\lambda}_3)-\bar{f}(0)) {\bar{\lambda}_1}^2 \} \\
&=& \bar{f}'(0) - \dfrac{\bar{\lambda}_1 \bar{\lambda}_3 (T_2(\bar{\lambda}_1)-T_2(\bar{\lambda}_3)) }{\bar{\lambda}_1-\bar{\lambda}_3}, \\
\bar{c} &=& \bar{s}+\bar{t}+\bar{u} \\
&=& \frac{1}{\bar{\lambda}_1\bar{\lambda}_3(\bar{\lambda}_1-\bar{\lambda}_3)} 
\{ f(\bar{\lambda}_1) \bar{\lambda}_3 + \bar{f}(0)(\bar{\lambda}_1-\bar{\lambda}_3) - f(\bar{\lambda}_3) \bar{\lambda}_1 \}\\
&=& \frac{1}{\bar{\lambda}_1\bar{\lambda}_3(\bar{\lambda}_1-\bar{\lambda}_3)} 
\{ (f(\bar{\lambda}_1)-\bar{f}(0)) \bar{\lambda}_3  - (f(\bar{\lambda}_3)-\bar{f}(0)) \bar{\lambda}_1 \} \\
&=& \frac{\bar{\lambda}_1 T_2(\bar{\lambda}_1)-\bar{\lambda}_3 T_2(\bar{\lambda}_3)}{\bar{\lambda}_1-\bar{\lambda}_3}.
\end{eqnarray*}
Although $\bar{b}$ and $\bar{c}$ 
may suffer from small denominators when $\bar{\lambda}_1 \to \bar{\lambda}_3$, 
it can be stably computed 
by substituting $T_2(\lambda_1)$ with its Taylor expansion
$T_2(\lambda_1) = T_2(\lambda_3) + (\lambda_1-\lambda_3) 
T_3(\lambda_1-\lambda_3)$, where $T_3$ is an analytic function.
Notice that 
since we assumed $\lambda_1 \ge \lambda_2 \ge \lambda_3$, 
$|\bar{\lambda}_1-\bar{\lambda}_3| \to 0$ implies $\bar{\lambda}_1, \bar{\lambda}_3 \to 0$.
For $f=\exp$ we have
\[
T_2(x) =  \exp(\lambda_2)\dfrac{\exp(x)-1-x}{x^2} = \exp(\lambda_2) \left(\frac12 + \frac{x}{3!} + \frac{x^2}{4!}+\cdots \right)
\]
to obtain
\begin{eqnarray*}
 \bar{b} &=& \exp(\lambda_2)\left( 1 - \frac{\bar{\lambda}_1 \bar{\lambda}_3}{3!}+ \cdots \right)\\ 
 \bar{c} &=& \exp(\lambda_2)\left(  \frac12 + \frac{\bar{\lambda}_1 +\bar{\lambda}_3}{3!} + \frac{{\bar{\lambda}_1}^2 + \bar{\lambda}_1 \bar{\lambda}_3 +{\bar{\lambda}_3}^2}{4!} \cdots \right).
\end{eqnarray*}
These sequences converge quickly when $\bar{\lambda}_1, \bar{\lambda}_3 \to 0$.

To derive the formula for $\log$ in \S \ref{closed-formula-of-log},
we put $\bar{Y}=Y/\lambda_2$ and argue similarly using
$\log(\bar{Y})=\log(Y)-\log(\lambda_2)I_3$.

%
%

\section*{Acknowledgement}
 This research was conducted as a part of the 
 Core Research for Evolutional Science and Technology(CREST) Program 
``Mathematics for Expressive Image Synthesis'' of the Japan Science and Technology Agency (JST).
  We would like to thank the members of the project, especially, Ken Anjyo,
 Sampei Hirose, Kohei Matsushita, Yoshihiro Mizoguchi, Hideki Todo, and Shun'ichi Yokoyama
  for valuable discussions and comments.
  We also appreciate the referees for their careful reading and insightful comments.

\bibliographystyle{acm}
\bibliography{WEB-exponential}

\end{document}